\documentclass[twocolumn,showpacs,preprintnumbers,amsmath,amssymb]{revtex4}
\usepackage{amsthm}

\newcommand\R{\mathbb R}
\newcommand\Sym{\mathbb{S}}
\newcommand\T{\mathbb{T}}
\newtheorem{thm}{Theorem}
\newtheorem{lem}[thm]{Lemma}
\renewcommand{\d}{{\,\rm d}}

\begin{document}


\title{A new first-order formulation for the Einstein equations} 
\author{Alexander M.~Alekseenko}%
 \email{alekseen@math.umn.edu}
 \homepage{http://www.math.umn.edu/~alekseen}
 \affiliation{School of Mathematics, University of Minnesota, Minneapolis,
              Minnesota 55455}%
\author{Douglas N.~Arnold}
 \email{arnold@ima.umn.edu}
 \homepage{http://www.ima.umn.edu/~arnold}
 \affiliation{Institute for Mathematics and its Applications, University
             of Minnesota, Minneapolis, Minnesota 55455}   
 \thanks{The second author was supported by NSF grant DMS-0107233.}
\date{October 18, 2002}

\begin{abstract}
We derive a new first-order formulation for Einstein's equations which
involves fewer unknowns than other first-order formulations that have
been proposed. The new formulation is based on the $3+1$ decomposition
with arbitrary lapse and shift.  In the reduction to first order form
only 8 particular combinations of the 18 first derivatives of the spatial
metric are introduced.  In the case of linearization about Minkowski
space, the new formulation consists of symmetric hyperbolic system in
14 unknowns, namely the components of the extrinsic curvature
perturbation and the 8 new variables, from whose solution the metric
perturbation can be computed by integration.
\end{abstract}

\maketitle

\section{Introduction}

Many ways have been proposed to formulate Einstein's equations of
general relativity in a manner suitable for numeric computation. In
this paper we introduce a new first-order
formulation for Einstein's equations.  This system involves fewer
unknowns than other first-order formulations that have been proposed
and does not require any arbitrary parameters.  In the simplest case of
linearization around Minkowski space with constant lapse and
vanishing shift, the system has the simple form:
\begin{equation}\label{BASIC}
\frac1{\sqrt2}\partial_t \kappa_{ij} = \partial^l\lambda_{l(ij)},
\quad
\frac1{\sqrt2}\partial_t \lambda_{lji}=\partial_{[l}\kappa_{j]i}.
\end{equation}
Here $\kappa_{ij}$ is the extrinsic curvature perturbation, a symmetric
tensor, and $\lambda_{ijk}$ is a third-order tensor which is
antisymmetric with respect to the first two indices and satisfies
a cyclic identity, with the result that the system above
is symmetric hyperbolic in $14$ unknowns.

Our approach applies as well to the full nonlinear ADM system with
arbitrary lapse and shift.  We work with the actual lapse, rather than
a densitized version.  In the nonlinear case the system involve 20
unknowns and can be written
\begin{gather*}
\partial_0 h_{ij} = -2ak_{ij} + 2h_{s(i}\partial_{j)}b^{s},\\
\frac1{\sqrt2}\partial_0 k_{ij} = a h_{mi}h_{nj}\partial_l
  f^{l(mn)}+\cdots,\\
\frac1{\sqrt2}\partial_0 f_{lmn} = \partial_{[l}(ak)_{m]n} + \cdots.
\end{gather*}
Here $\partial_0:=\partial_t-b^l\partial_l$ is the convective
derivative, indices are raised and lowered using the spatial metric
components $h_{ij}$ and the omitted terms are algebraic expressions
in the $h_{ij}$, their spatial derivatives
$\partial_lh_{ij}$, and the extrinsic curvature components $k_{ij}$, and also involve
the lapse $a$ and shift components $b^i$.  The $f_{lmn}$, which depend on the first
spatial derivatives of the spatial metric satisfy the same symmetries
as in the linear case, and so represent $8$ unknowns.

As is common, our derivation will start from the Arnowitt--Deser--Misner
$3+1$ decomposition \cite{ADM}; see also \cite{York}. The ADM approach introduces
a system of coordinates $t=x_0$, $x_1$, $x_2$, $x_3$, with
$t$ a timelike variable and the $x_i$ spacelike for $i=1,2,3$, and encodes the
4-metric of spacetime as a time-varying 3-metric on a 3-dimensional
domain together with the lapse and
shift, which are scalar-valued and 3-vector-valued functions of time and
space, respectively.
Specifically the coordinates of the 4-metric are given by
\begin{equation*}
g_{00}=-a^2+b_ib_jh^{ij},\quad
g_{0i}=b_i,\quad
g_{ij}=h_{ij}.
\end{equation*}
Here $a$ denotes the lapse, the $b_i$ are the components of the
shift vector $b$, and the $h_{ij}$ the components of the spatial
metric $h$.  As usual Roman indices run from $1$ to $3$ and
$(h^{ij})$ denotes the matrix inverse to $(h_{ij})$. Let $D_i$
denote the covariant derivative operator
associated to the spatial metric and set
\begin{equation*}
k_{ij}=-\frac{1}{2a}\partial_{t}h_{ij}+\frac{1}{a}D_{(i}b_{j)},
\end{equation*}
the extrinsic curvature.  Then the ADM equations for a vacuum spacetime
are
\begin{gather}
\label{ADM1}
\partial_{t} h_{ij}
= -2 a k_{ij}+2D_{(i}b_{j)},
 \\
\partial_{t} k_{ij}
= a[R_{ij} + (k_l^l) k_{ij}-2k_{il}k^{l}_{j}] + b^{l}D_{l}k_{ij} \nonumber \\
\label{ADM2}
  \hspace{6mm}{}+k_{il}D_{j}b^{l}+k_{lj}D_{i}b^{l}-D_{i}D_{j}a, \\
\label{ADM3}
R_i^i+(k_i^i)^2 - k_{ij}k^{ij} =0,
\\
\label{ADM4}
D^j k_{ij}-D_i k_j^j = 0.
\end{gather}
Here $R$ denotes the spatial Ricci tensor, whose components are given
by second order spatial partial differential
operators applied to the spatial metric components.  Also indices are raised
and traces taken with respect to the spatial metric.
We have used the
notation of parenthesized indices to denote the symmetric part
of a tensor:
$D_{(i}b_{j)}:=(D_i b_j+D_j b_i)/2$.
Later we will similarly use bracketed indices
to denote the antisymmetric part and sometimes bars to separate
indices, so, as an example,
$u_{[i|j|k]l}:=(u_{ijkl}-u_{kjil})/2$.

Equations (\ref{ADM3}) and (\ref{ADM4}), called the Hamiltonian and momentum constraints,
do not involve time differentiation. The first two equations
are the evolution equations. A typical approach is to determine
the lapse and shift in some way, find relevant initial data for
$h$ and $k$ satisfying the constraint equations, and then to
solve the evolution equations to determine the metric and extrinisic
curvature for future times. The constraint equations may or may
not be explicitly imposed during the evolution. For exact
solutions of the evolution equation with initial data exactly
satisfying the constraints, the constraints are automatically
satisfied for future times.

The system of evolution equations for $h$ and $k$ is first order
in time and second order in space.  They are not hyperbolic in
any usual sense, and their direct discretization seems
difficult. Therefore, many authors \cite{FR, AbY, AnY, KST}
have considered reformulations into more standard first order
hyperbolic systems. Typically these approaches involve
introducing all the first spatial derivatives of the spatial
metric components, or quantities closely related to them, as 18
additional  unknowns.  The resulting systems
involve many variables, sometimes 30 or more.  In the formulation
proposed here, we introduce only 8 particular combinations of the
first derivatives of the metric components.

In the next section of the paper we present our approach as applied to
a linearization of the ADM system.  This allows us to demonstrate the
basic ideas with a minimal of technical complications and to rigorously
establish the relationship between the new formulation and the ADM
system. In the third and final section of the paper we carry out the
derivation in the case of the full nonlinear ADM system.

\section{Symmetric formulation for the linearized system}\label{sec2}

We linearize the ADM equations about the trivial solution
obtained by representing
Minkowski spacetime in Cartesian coordinates: $h_{ij}=\delta_{ij}$, $k_{ij}=0$,
$a=1$, $b_i=0$.  Consider a perturbation given by
$h_{ij}=\delta_{ij}+\gamma_{ij}$, $k_{ij}=\kappa_{ij}$, $a=1+\alpha$,
$b_i=\beta_i$, with the $\gamma_{ij}$, $\kappa_{ij}$, $\alpha$, and
$\beta_i$ supposed to be small.  Substituting these expressions
into the ADM system and ignoring terms which are at least quadratic
with respect to the $\gamma_{ij}$, $\kappa_{ij}$, $\alpha$, and
$\beta_i$ and their derivatives, we obtain the linear system
we shall study:
\begin{gather}
\label{LADM1}
\partial_{t}\gamma_{ij}
= -2 \kappa_{ij}+2\partial_{(i}\beta_{j)}, \\
\label{LADM2}
\partial_{t}\kappa_{ij}=(P\gamma)_{ij} -\partial_{i}\partial_{j}\alpha, \\
\label{LADM3}
(P\gamma)^{i}_{i} =0,
\\
\label{LADM4}
\partial^j\kappa_{ij}-\partial_i \kappa^{j}_{j} = 0.
\end{gather}
Here $P\gamma$ is the linearized Ricci tensor, with components given by
\begin{equation}\label{DEFP}
(P\gamma)_{ij}=\frac{1}{2} \partial_{i}\partial^{l}\gamma_{lj}+
\frac{1}{2}\partial_{j}\partial^{l}\gamma_{li}-
\frac{1}{2}\partial^{l}\partial_{l}\gamma_{ij}-
\frac{1}{2}\partial_{i}\partial_{j}\gamma^{l}_{l}.
\end{equation}
In these expressions, and in general when we deal with the linearized
formulation, indices are raised and lowered with respect to
the Euclidean metric in $\R^3$, so, for example, $\partial^i$ and
$\partial_i$ are identical.

In order to reduce the linearized ADM system to first-order symmetric
hyperbolic form, we first develop an identity for $P\gamma$ valid
for any symmetric matrix field $\gamma$.  It is useful to introduce
the notations
$\Sym$ for the the 6-dimensional space of
symmetric matrices and $\T$ for the 8-dimensional space of triply-indexed
arrays $(w_{ijk})$ which are skew symmetric in the first two indices and
satisfy the cyclic property $w_{ijk}+w_{jki}+w_{kij}=0$.

We define the operator $M: C^{\infty}(\R^{3},\Sym)\to
C^{\infty}(\R^{3},\R^3)$ by
$(Mu)_{i}=\partial^{l}u_{il}-\partial_{i}u^{l}_{l}$.  (Note that the
linearized momentum
constraint, \eqref{LADM4}, is simply $M\kappa=0$.) From the definition
\eqref{DEFP}, we have
\begin{equation}\label{LP1}
(P\gamma)_{ij}=-\partial^l\partial_{[l}\gamma_{j]i}+\frac12\partial_i(M\gamma)_
j.
\end{equation}
Now, for any vector-valued function $v$, we have
\begin{equation*}
\partial_iv_j =
\partial^lv_j\delta_{li}=2\partial^lv_{[j}\delta_{l]i}+\partial^lv_l\delta_{ij}
.
\end{equation*}
Applying this identity to \eqref{LP1} with $v=M\gamma$ we get
\begin{equation}\label{LP2}
(P\gamma)_{ij}=-\partial^l[\partial_{[l}\gamma_{j]i}+(M\gamma)_{[l}\delta_{j]i}
]
 +\frac12\partial^l(M\gamma)_l\delta_{ij}.
\end{equation}

Define operators
$L:C^{\infty}(\R^{3}, \Sym)\to C^{\infty}(\R^{3}, \T)$, and
$L^{\ast}:C^{\infty}(\R^{3}, \T)\to C^{\infty}(\R^{3}, \Sym)$ by
\begin{equation*}
(Lu)_{lji}=\partial_{[l}u_{j]i},\quad
(L^{\ast}w)_{ij}=-\partial^{l}w_{l(ij)}.
\end{equation*}
One easily verifies that operators $L$ and $L^{\ast}$ are formal
adjoints to each other with respect to the scalar products
$\langle u,v \rangle=\int u_{pq}v^{pq}\d x$
and $\langle z,w \rangle=\int z_{pqr}w^{pqr} \d x$ in the
spaces $C^{\infty}(\R^3, \Sym)$ and $C^{\infty}(\R^3, \T)$
respectively.  Introducing
\begin{equation}\label{LAMBDA}
\lambda_{lji}=-\frac1{\sqrt2}[(L\gamma)_{lji}+(M\gamma)_{[l}\delta_{j]i}]
\end{equation}
we can then restate \eqref{LP2} as
\begin{equation*}
(P\gamma)_{ij}=\sqrt2\partial^l\lambda_{lji}
 +\frac12\partial^l(M\gamma)_l\delta_{ij}.
\end{equation*}
(The reason for the factor of $\sqrt 2$ will become apparent shortly.)
Taking symmetric parts, the last equation becomes
\begin{equation}\label{LID}
(P\gamma)_{ij}=-\sqrt2(L^*\lambda)_{ij}
 +\frac12\partial^l(M\gamma)_l\delta_{ij}.
\end{equation}

We are now ready to introduce our first-order symmetric hyperbolic
formulation.  The unknowns will be $\kappa\in C^\infty(\R^3,\Sym)$
and $\lambda\in C^\infty(\R^3,\T)$, so the system has $14$ independent
variables in all.  Substituting \eqref{LID} in \eqref{LADM2} and
noting that $\partial^l(M\gamma)_l=(P\gamma)_l^l=0$ by the linearized
Hamiltonian constraint
\eqref{LADM3}, we obtain an evolution equation for $\kappa$:
\begin{equation}\label{EKAPPA}
\partial_t\kappa_{ij}=-\sqrt2(L^*\lambda)_{ij}-\partial_i\partial_j\alpha.
\end{equation}
To obtain an evolution equation for $\lambda$, we differentiate
\eqref{LAMBDA} with respect to time and substitute \eqref{LADM1} to
eliminate~$\gamma$.  Simplifying and using the linearized momentum
constraint \eqref{LADM4} we obtain
\begin{equation}\label{ELAMBDA}
\partial_t\lambda_{lji}=\sqrt2(L\kappa)_{lji}-\tau_{lji},
\end{equation}
where $\tau$ can be determined from
$(\epsilon\beta)_{ij}:=\partial_{(i}\beta_{j)}$ by
\begin{eqnarray*}
\tau_{lji}
&=&\sqrt2[(L\epsilon\beta)_{lji}+(M\epsilon\beta)_{[l}\delta_{j]i}] \\
&=&\frac1{\sqrt2}(\partial_i\partial_{[l}\beta_{j]}
   +\partial^m\partial_{[m}\beta_{l]}\delta_{ij}
   -\partial^m\partial_{[m}\beta_{j]}\delta_{il}).
\end{eqnarray*}
Equations \eqref{EKAPPA} and \eqref{ELAMBDA} constitute a first-order
symmetric hyperbolic system (this is clear, since $L$ and $L^*$ are
formal adjoints).
It follows that (see, e.g., \cite{EVA} \S~7.3.2),
if the laspe and the shift are sufficiently smooth then for
arbitrary initial values $\kappa(0)$ and $\lambda(0)$
belonging to $H^1(\R^3)$, there exists a unique solution to system
\eqref{EKAPPA} and \eqref{ELAMBDA} with components in
$H^1((0,T)\times\R^3)$.

The Cauchy problem for the original linearized ADM system consists
of the equations \eqref{LADM1}--\eqref{LADM4} together with
specific initial values $\gamma(0)$ and $\kappa(0)$.
The foregoing derivation shows that if $\gamma$ and $\kappa$ satisfy
the ADM system and $\lambda$ is defined by (\ref{LAMBDA}), then
$\kappa$ and $\lambda$ satisfy the symmetric
hyperbolic system (\ref{EKAPPA}), (\ref{ELAMBDA}).
Conversely, to recover the solution to the ADM system from
(\ref{EKAPPA}), (\ref{ELAMBDA}), the same initial condition should be
imposed on $\kappa$ and $\lambda$ should  be taken initially to be
\begin{equation}\label{ILAMBDA}
\lambda_{lji}(0)=
-\frac1{\sqrt2}[(L\gamma(0))_{lji}+(M\gamma(0))_{[l}\delta_{j]i}].
\end{equation}
Once $\kappa$ and $\lambda$ are determined, the metric perturbation
$\gamma$ is given by
\begin{equation}\label{GAMMA}
\gamma_{ij} = \gamma_{ij}(0)-2\int_0^t
(\kappa_{ij}-\partial_{(i}\beta_{j)}),
\end{equation}
as follows from \eqref{LADM1}.

\begin{thm}
Let the lapse perturbation $\alpha$ and shift perturbation $\beta$ be
given.  Suppose that initial data $\gamma(0)$ and $\kappa(0)$ are
specified satisfying the constraint equations \eqref{LADM3},
\eqref{LADM4} at time $t=0$.  Define $\lambda(0)$ by \eqref{ILAMBDA}, and
determine
$\kappa$ and $\lambda$ from the first-order symmetric hyperbolic
system \eqref{EKAPPA}, \eqref{ELAMBDA}.  Finally, define $\gamma$
by \eqref{GAMMA}.  Then the ADM system \eqref{LADM1}--\eqref{LADM4}
is satisfied.
\end{thm}
\begin{proof}
Equation \eqref{LADM1} follows from \eqref{GAMMA} by differentiation.

Next we verify the momentum constraint \eqref{LADM4}.  To do so we
will show that $\mu:=M\kappa$ satisfies a second order wave equation,
and that $\mu(0)=\partial_t\mu(0)=0$.
Indeed $\mu(0)=0$ by assumptions.  To see that $\partial_t\mu(0)=0$,
we apply the operator $M$ to \eqref{EKAPPA} and use the fact that
$M$
annihilates the Hessian $\partial_i\partial_j\alpha$ for any function
$\alpha$.  Therefore $\partial_t\mu=-\sqrt2ML^*\lambda$.  Using \eqref{LID}
and the assumption that $\gamma$ satisfies the Hamiltonian constraint
at the initial time, we find that
\begin{equation*}
\partial_t\mu_i(0)=(MP\gamma(0))_i=-\frac{1}{2}\partial_i\partial^l(M\gamma(0))_l=0.
\end{equation*}

To obtain a second order equation for $\mu$, first we
differentiate \eqref{EKAPPA} in time and substitute \eqref{ELAMBDA}
to get a second-order equation for $\kappa$:
\begin{equation*}
\partial_t^2\kappa_{ij}=-2(L^*L\kappa_{ij})-\partial_i\partial_j\alpha.
\end{equation*}
Here we have used the fact that $L^*\tau\equiv 0$.  Apply $M$ to
the last equation.  Using the identity $(ML^*L\kappa)=
-\partial^l\partial_{(i}(M\kappa)_{l)}/2$ and the fact that $M$
annihilates Hessians, we find that $\mu:=M\kappa$ satisfies the second-order
hyperbolic equation
\begin{equation*}
\partial_t^2 \mu_i = \partial^l\partial_{(i}\mu_{l)}.
\end{equation*}
This is simply an elastic wave equation.  Since
$\mu(0)=\partial_t\mu(0)=0$, $\mu$ vanishes for all time, i.e., the
momentum constraint is satisfied.

Now $(P\kappa)_i^i=\partial^i(M\kappa)_i=0$. Moreover, $P$ applied
to $\epsilon\beta$ is identically zero.  Therefore if we apply
$P$ to \eqref{GAMMA} and take the trace, we find that
$(P\gamma)_i^i=(P\gamma(0))_i^i$, which vanishes by assumption.
This verifies the Hamiltonian constraint \eqref{LADM3}.

It remains to verify \eqref{LADM2} which, in view of \eqref{EKAPPA}
comes down to showing that $\sqrt2L^*\lambda=-P\gamma$.  Since we
have verified the Hamiltonian constraint, this will follow if we
can establish \eqref{LID}, which is itself a consequence of
\eqref{LAMBDA}.  We used \eqref{LAMBDA} at time $t=0$ to initialize
$\lambda$, so it is sufficient to show that
\begin{equation*}
\partial_t\lambda_{lji}=-\frac1{\sqrt2}\partial_t[(L\gamma)_{lji}+(M\gamma)_{[l
}\delta_{j]i}].
\end{equation*}
This follows directly from \eqref{ELAMBDA} and \eqref{LADM1}.
\end{proof}

We conclude this section by computing the plane wave solutions
to the hyperbolic system \eqref{BASIC}.  That is, we seek solutions
of the form
\begin{equation*}
\kappa_{ij}=\tilde{\kappa}_{ij}f(st-n^{a}x_{a}),\qquad
\lambda_{lji}=\tilde{\lambda}_{lji}f(st-n^{a}x_{a}),
\end{equation*}
where the real number $s$ gives the wave speed and the unit vector $n_{i}$ the
wave direction, the polarizations $\tilde{\kappa}_{ij}$ and $\tilde{\lambda}_{lji}$ are
constant, and the profile $f(t)$ is an
arbitrary differentiable function.

Substituting these expressions into \eqref{BASIC}, we get
\begin{gather*}
s\tilde{\kappa}_{ij}f'(st-n^{a}x_{a}) = -\sqrt2 n^{l}\tilde{\lambda}_{l(ij)}f'(st-n^{a}x_{a}),
\\
s\tilde{\lambda}_{lji}f'(st-n^{a}x_{a}) = -\sqrt2 n_{[l}\tilde{\kappa}_{j]i}f'(st-n^{a}x_{a}),
\end{gather*}
and so the system reduces to the following linear eigenvalue problem:
\begin{gather}
\label{EV1}
s\tilde{\kappa}_{ij}=-\sqrt{2}n^{l}\tilde{\lambda}_{l(ij)},\\
\label{EV2}
s\tilde{\lambda}_{lji}=-\sqrt{2}n_{[l}\tilde{\kappa}_{j]i},
\end{gather}
with the wave speed $s$ as eigenvalues and the pairs
$(\tilde{\kappa}_{ij}, \tilde{\lambda}_{lji})$ as eigenvectors.
The eigenvalues of this system are $0$ (multiplicity 4), $\pm 1$
(each multiplicity $3$), and $\pm1/\sqrt2$ (each multiplicity 2).
To verify this and describe the eigenvectors we introduce a
unit vector
$m_{i}$ be a perpendicular to $n_i$, and set
$l_{i}=\varepsilon_{i}^{\phantom{i}ab}n_{a}m_{b}$ to
complete an orthonormal frame.  Then the following solution to
the eigenvalue problem can be
checked by direct substitution into \eqref{EV1}--\eqref{EV2}:
\begin{align*}
&\text{$s=0$: }&
&(0,m_{[l}l_{j]}m_{i}),\quad
(0,m_{[l}l_{j]}l_{i}),\quad
(n_{i}n_{j},0),
\\
&&&(0,2m_{[l}l_{j]}n_{i}+n_{[l}l_{j]}m_{i}-n_{[l}m_{j]}l_{i});
\\
&\text{$s=\pm1$: }&
&(l_{i}l_{j}-m_{i}m_{j},\mp\sqrt{2}(n_{[l}l_{j]}l_{i}-n_{[l}m_{j]}m_{i})),
\\
&&&(l_{i}m_{j}+m_{i}l_{j},\mp\sqrt{2}(n_{[l}l_{j]}m_{i}+n_{[l}m_{j]}l_{i})),
\\
&&&(l_{i}l_{j}+m_{i}m_{j},\mp\sqrt{2}(n_{[l}l_{j]}l_{i}+n_{[l}m_{j]}m_{i}));
\\
&\text{$s=\pm1/\sqrt{2}$: }&
& (n_{(i}l_{j)}, \mp\,n_{[l}l_{j]}n_{i}),
\\
&&&(n_{(i}m_{j)}, \mp\,n_{[l}m_{j]}n_{i}).
\end{align*}

\section{Decomposition of the full ADM system}

In this section we develop a first order formulation of the full
nonlinear ADM system analogous to that developed for the linearized
system in Section~\ref{sec2}. We continue to assume that the underlying
manifold is topologically $\R^3$ and view the ADM system as equations
for the evolution of a Riemannian $3$-metric $h$ on $\R^3$.
Thus $h_{ij}$, $1\le i,j\le 3$ are the components of
the spatial metric.  They form a positive-definite symmetric matrix
defined at each point of $\R^3$ and varying in time.  Indices on
other fields are lowered and raised using $h_{ij}$ and the inverse matrix
field $h^{ij}$.

For the components of the Riemann tensor we have
\begin{equation*}
R_{ijkl}=\partial_j\Gamma_{ikl}-\partial_i\Gamma_{jkl}+
h^{mn}(\Gamma_{jkm}\Gamma_{iln}-\Gamma_{ikm}\Gamma_{jln}),
\end{equation*}
where the Cristoffel symbols are defined by
$\Gamma_{ijk}=(\partial_{i}h_{jk}+\partial_{j}h_{ki}-\partial_{k}h_{ij})/2$.
The components of the Ricci tensor are given by $R_{ij}=h^{pq}R_{piqj}$,
which yields
\begin{eqnarray*}
R_{ij}
&=&\frac{1}{2}h^{pq}(\partial_{p}\partial_{j}h_{iq}+
   \partial_{i}\partial_{p}h_{qj}-\partial_{p}\partial_{q}h_{ij}-
   \partial_{i}\partial_{j}h_{pq}) \\
& &{}+h^{pq}h^{rs}(\Gamma_{ips}\Gamma_{qjr}-\Gamma_{pqs}\Gamma_{ijr}).
\end{eqnarray*}

We define a second-order linear partial differential operator
$P:C^\infty(\R^3,\R^{3\times3})\to C^\infty(\R^3,\R^{3\times3})$ by
\begin{eqnarray*}
(Pu)_{ij}
&=&\frac{1}{2}[
  \partial_{p}(h^{pq}\partial_{j}u_{qi})
+ \partial_{i}(h^{pq}\partial_{p}u_{jq}) \nonumber \\
& &\hspace{5mm} {}- \partial_{p}(h^{pq}\partial_{q}u_{ji})
- \partial_{i}(h^{pq}\partial_{j}u_{pq})].\qquad
\end{eqnarray*}
Note that in the case $h_{ij}=\delta_{ij}$, $P$ coincides with the
linearized Ricci operator introduced in (\ref{DEFP}).  In
the nonlinear case, $P$ is related to the Ricci tensor by the
equation
\begin{equation}
\label{R1}
R_{ij}=(Ph)_{ij}+c^{1)}_{ij},
\end{equation}
where
\begin{eqnarray*}
c^{1)}_{ij}
&:=&-\frac{1}{2}[(\partial_{p}h^{pq})(\partial_{j}h_{iq})+
     (\partial_{i}h^{pq})(\partial_{p}h_{qj}) \\
& & \hspace{1cm}{}-(\partial_{p}h^{pq})(\partial_{q}h_{ij})-
    (\partial_{i}h^{pq})(\partial_{j}h_{pq})] \\
& &{}+h^{pq}h^{rs}(\Gamma_{ips}\Gamma_{qjr}-\Gamma_{pqs}\Gamma_{ijr}),
\end{eqnarray*}
is a first-order differential operator in $h_{ij}$.

Define an operator $M:C^{\infty}(\R^{3},\Sym) \to C^{\infty}(\R^{3},\R^{3})$ by
\begin{equation*}
(Mu)_{i} = 2h^{pq}\partial_{[p}u_{i]q}.
\end{equation*}
This formula extends the definition of the linearized momentum constraint
operator introduced in the previous section.
Up to lower order terms the momentum constraint (\ref{ADM4})
is given by the vanishing of $Mk$:
\begin{equation}
\label{M2}
D^{p}k_{ip}-D_{i}k^{p}_{p} = 2h^{pq}D_{[p}k_{i]q}=(Mk)_{i} -
2h^{pq}\Gamma^{r}_{q[p} k^{\hphantom{r}}_{i]r}
\end{equation}
(where $\Gamma^r_{qp}:=h^{rs}\Gamma_{qps}$).

Finally, we introduce operators $L:C^{\infty}(\R^3,\Sym) \to C^{\infty}(\R^{3},\T)$
and $L^{\ast}:C^{\infty}(\R^{3},\T) \to C^{\infty}(\R^3,\Sym)$ by
\begin{equation*}
(Lu)_{ijl}=\partial_{[i}u_{j]l},\qquad (L^{\ast}v)_{ij}=-h_{qi}h_{rj}\partial_{
p}v^{p(qr)}.
\end{equation*}
As in linear case, the operators $L$ and $L^{\ast}$ are formal
adjoints to each other with respect to the scalar products
$\langle u,w \rangle=\int u_{ij}w^{ij}\d x$ and
$\langle v,z \rangle=\int v_{ijl}z^{ijl}\d x$ on the spaces
$C^{\infty}(\R^3,\Sym)$ and
$C^{\infty}(\R^{3},\T)$ respectively.

Finally, we introduce new variables
\begin{equation}
\label{F1}
f_{lmn}=-\frac{1}{\sqrt{2}}[(Lh)_{lmn}+(Mh)_{[l}h_{m]n}],
\end{equation}
and develop the analogue of the identity (\ref{LID}).
\begin{lem}
The following identity is valid for the Ricci tensor
\begin{equation}
\label{R2}
R_{ij}=-\sqrt{2}(L^{\ast}f)_{ij} + \frac{1}{2}\partial_{p}[h^{pq}(Mh)_{q}h_{ij}
] + c^{2)}_{ij},
\end{equation}
where
\begin{multline*}
c^{2)}_{ij} = c^{1)}_{(ij)}+
h_{qi}h_{rj}[\partial_{p}(h^{qm}h^{rn})]h^{pl}[(Lh)_{l(mn)} \\
+\frac{1}{2}(Mh)_{[l}h_{m]n}+\frac{1}{2}(Mh)_{[l}h_{n]m}]
\end{multline*}
is first order in $h_{ij}$.
\end{lem}
\begin{proof}
The formula (\ref{R2}) is a consequence of the identity
\begin{equation}
\label{L2}
-\sqrt{2}(L^{\ast}f)_{ij}
=(Ph)_{(ij)}-\frac{1}{2}\partial_{p}[h^{pq}(Mh)_{q}h_{ij}]+c^{3)}_{ij},
\end{equation}
where
\begin{multline*}
c^{3)}_{ij}=-h_{qi}h_{rj}[\partial_{p}(h^{qm}h^{rn})]h^{pl}[(Lh)_{l(mn)} \\
 +\frac{1}{2}(Mh)_{[l}h_{m]n}+\frac{1}{2}(Mh)_{[l}h_{n]m}].
\end{multline*}
To prove the identity (\ref{L2}) we note, first,
that operator $P$ can be rewritten in terms of operators $L$ and $M$ as
\begin{equation*}
(Pu)_{ij}=-\partial_{p}[h^{pq}(Lu)_{qji}]+\frac{1}{2}\partial_{i}(Mu)_{j}
\end{equation*}
which yields
\begin{equation*}
(Pu)_{(ij)}=-\partial_{p}[h^{pq}(Lu)_{q(ij)}]+\frac{1}{2}\partial_{(i}(Mu)_{j)},
\end{equation*}
and, second, that according to the definition of $L^{\ast}$,
\begin{eqnarray}
\label{L3}
(L^{\ast}v)_{ij}
&=&-h_{qi}h_{rj}\partial_{p}(h^{pl}h^{qm}h^{rn}v_{l(mn)}) \nonumber \\
&=&-\partial_{p}(h^{pq}v_{q(ij)}) \nonumber \\
& &   {}-h_{qi}h_{rj}[\partial_{p}(h^{qm}h^{rn})]h^{pl}v_{l(mn)}.
\end{eqnarray}
To derive the identity (\ref{L2}) we substitute $-\sqrt{2}f_{qij}$ for
$v_{qij}$ in (\ref{L3}). The first term on the right-hand side of (\ref{L3})
then becomes
\begin{eqnarray*}
\lefteqn{-\partial_{p}(h^{pq}[(Lh)_{q(ij)}+\frac{1}{2}(Mh)_{q}h_{ij}
-\frac{1}{2}(Mh)_{(i}h_{j)q}]) }\quad
& & \\
&=& - \partial_{p}[h^{pq}(Lh)_{q(ij)}]
    - \frac{1}{2}\partial_{p}[h^{pq}(Mh)_{q}h_{ij}] \\
& & \hspace{5mm}{}+ \frac{1}{2}\partial_{p}[h^{pq}(Mh)_{(i}h_{j)q}] \\
&=& - \partial_{p}[h^{pq}(Lh)_{q(ij)}]
    - \frac{1}{2}\partial_{p}[h^{pq}(Mh)_{q}h_{ij}] \\
& & \hspace{5mm}{}+ \frac{1}{2}\partial_{(i}(Mh)_{j)} \\
&=& (Ph)_{(ij)}-\frac{1}{2}\partial_{p}[h^{pq}(Mh)_{q}h_{ij}].
\end{eqnarray*}
The substitution of $-\sqrt{2}f_{lmn}$ for $\nu_{lmn}$ into the second term of
the equation (\ref{L3}) gives the term $c^{3)}_{ij}$ precisely. The
rest of the proof follows from the identity~(\ref{R1}).
\end{proof}

We now proceed to the derivation of the new formulation of the ADM system.
In (\ref{ADM2}) we substitute $\partial_0+b^{l}\partial_{l}$ for
$\partial_t$ and
replace $R_{ij}$ with the right side of the
(\ref{R2}) to get
\begin{equation}
\label{K1}
\partial_{0}k_{ij}=
-\sqrt{2}a(L^{\ast}f)_{ij}+\frac{1}{2}a\partial_{p}[h^{pq}(Mh)_{q}h_{ij}]+c^{4)
}_{ij},
\end{equation}
where
$$
c^{4)}_{ij}=a[c^{2)}_{ij}+(k^{l}_{l})k_{ij}-2k_{il}k^{l}_{j}]
+k_{il}\partial_{j}b^{l} + k_{lj}\partial_{i}b^{l} -D_{i}D_{j}a.
$$
Here we used the fact that the Lie derivative
$b^{l}D_{l}k_{ij}+k_{il}D_{j}b^{l}+k_{lj}D_{i}b^{l}
=b^{l}\partial_{l}k_{ij}+k_{il}\partial_{j}b^{l}+k_{lj}\partial_{i}b^{l}$.
We treat the second term on the right-hand side of (\ref{R2}) using
the Hamiltonian constraint (\ref{ADM3}).
Now,
\begin{eqnarray*}
R^{i}_{i}
&=& \frac{1}{2}h^{ij}h^{pq}(\partial_{p}\partial_{j}h_{iq}+
    \partial_{i}\partial_{p}h_{qj}-\partial_{p}\partial_{q}h_{ij}-
    \partial_{i}\partial_{j}h_{pq}) \\
& &\hspace{1mm}{}+h^{ij}h^{pq}h^{rs}
  (\Gamma_{ips}\Gamma_{qjr}-\Gamma_{pqs}\Gamma_{ijr}) \\
&=& h^{ij}h^{pq}(\partial_{i}\partial_{p}h_{qj}- \partial_{i}\partial_{j}h_{pq}) \\
& &\hspace{1mm}{}+h^{ij}h^{pq}h^{rs}
  (\Gamma_{ips}\Gamma_{qjr}-\Gamma_{pqs}\Gamma_{ijr}) \\
&=& \partial_{i}[h^{ij}h^{pq}(\partial_{p}h_{qj}-\partial_{j}h_{pq})] \\
& &\hspace{1mm}{}-[\partial_{i}(h^{ij}h^{pq})](\partial_{p}h_{qj}-\partial_{j}h_{pq})\\
& &\hspace{1mm}{}+h^{ij}h^{pq}h^{rs}
  (\Gamma_{ips}\Gamma_{qjr}-\Gamma_{pqs}\Gamma_{ijr})\\
&=& \partial_{i}[h^{ij}(Mh)_{j}] + c^{5)},
\end{eqnarray*}
where
\begin{multline*}
c^{5)}=-[\partial_{i}(h^{ij}h^{pq})](\partial_{p}h_{qj}
-\partial_{j}h_{pq})\\
+h^{ij}h^{pq}h^{rs}
  (\Gamma_{ips}\Gamma_{qjr}-\Gamma_{pqs}\Gamma_{ijr}).
\end{multline*}
Hence, due to the Hamiltonian constraint,
\begin{eqnarray*}
\lefteqn{\partial_{p}[h^{pq}(Mh)_{q}h_{ij}]}\quad
& & \\
&=&\{\partial_{p}[h^{pq}(Mh)_{q}]\}h_{ij}+
   h^{pq}(Mh)_{q}(\partial_{p}h_{ij}) \\
&=& [k_{pq}k^{pq}-(k_p^p)^2-c^{5)}]h_{ij}+h^{pq}(Mh)_{q}(\partial_{p}h_{ij}).
\end{eqnarray*}
Combining all lower order terms into an expression $B_{ij}$,
first order in $h$,
we reduce (\ref{K1})  to
\begin{equation}
\label{K2}
\partial_{0}k_{ij}=-\sqrt{2}a(L^{\ast}f)_{ij}+B_{ij}.
\end{equation}
This is the first evolution equation of our system.

The second evolution equation will be obtained by applying
$\partial_{0}$ to the definition of $f$ (\ref{F1}):
\begin{equation*}
\partial_{0}f_{lmn}=-\frac{1}{\sqrt{2}}\{\partial_{0}(Lh)_{lmn}+\partial_{0}[(Mh)
_{[l}h_{m]n}]\}.
\end{equation*}
First, we note that
\begin{multline*}
\partial_{0}(Lh)_{lmn}
=(L\partial_{0}h)_{lmn} \\
+\frac{1}{2}[(\partial_{l}b^{s})(\partial_{s}h_{mn})
     -(\partial_{m}b^{s})(\partial_{s}h_{ln})].
\end{multline*}
Using the fact that $(Mu)_{l}=2h^{pq}(Lu)_{plq}$, we then get
\begin{multline*}
\partial_{0}(Mh)_{l}
= (M\partial_{0}h)_{l}+2(\partial_{0}h^{pq})(Lh)_{plq} \\
+h^{pq}[(\partial_{p}b^{s})(\partial_{s}h_{lq})
   -(\partial_{l}b^{s})(\partial_{s}h_{pq})] .
\end{multline*}
If we use this formula to compute $\partial_0[(Mh)_l h_{mn}]$ and
then antisymmetrize in $l$ and $m$,
we obtain
\begin{align}
\partial_{0}f_{lmn}
= &-\frac{1}{\sqrt{2}}[(L\partial_{0}h)_{lmn}+(M\partial_{0}h)_{[l}h_{m]n}]
\nonumber \\
&-\sqrt{2}(\partial_{0}h^{pq})(Lh)_{p[l|q|}h_{m]n}
\nonumber \\
&-\frac{1}{\sqrt{2}}(Mh)_{[l}\partial_{0}h_{m]n}+c^{6)}_{lmn},
\label{F3}
\end{align}
where
\begin{multline*}
c^{6)}_{lmn}
=-\frac{1}{2\sqrt{2}}[(\partial_{l}b^{s})(\partial_{s}h_{mn})
     -(\partial_{m}b^{s})(\partial_{s}h_{ln})]\\
-\frac{1}{\sqrt{2}}h^{pq}[(\partial_{p}b^{s})(\partial_{s}h_{q[l})h_{m]n}
   -(\partial_{s}h_{pq})(\partial_{[l}b^{s})h_{m]n}].
\end{multline*}

Next we use (\ref{ADM1}) to relate the terms in (\ref{F3})
involving $\partial_0 h$ to the extrinsic curvature $k$.
For the Lie derivative of the metric, we have
\begin{equation*}
2D_{(i}b_{j)} =b^s\partial_sh_{ij}+2h_{s(i}\partial_{j)}b^s.
\end{equation*}
Using this (\ref{ADM1}) becomes
\begin{equation}\label{ODE}
\partial_0 h_{ij} = -2ak_{ij} + 2w_{ij},
\end{equation}
where $w_{ij}:=h_{s(i}\partial_{j)}b^s$.
Using the Leibniz rule we can then verify that
\begin{equation*}
\partial_0 h^{ij} = 2ak^{ij} - 2w^{ij}.
\end{equation*}
Substituting these expressions in (\ref{F3}) we obtain
\begin{equation}
\label{F4}
\partial_{0}f_{lmn}=
\sqrt{2}[L(ak)]_{lmn}+\sqrt{2}[M(ak)]_{[l}h_{m]n}+c^{7)}_{lmn},
\end{equation}
where
\begin{eqnarray*}
c^{7)}_{lmn}
&=& c^{6)}_{lmn}-\sqrt{2}[(Lw)_{lmn} + (Mw)_{[l}h_{m]n}] \\
& &{}-2\sqrt{2}[ak^{pq}-w^{pq}](Lh)_{p[l|q|}h_{m]n} \\
& &{}+\sqrt2(Mh)_{[l}ak_{m]n}-\sqrt2(Mh)_{[l}w_{m]n}.
\end{eqnarray*}
The final step is to invoke the momentum constraint to simplify the
second term
on the right-hand side of (\ref{F4}). Indeed, since the
right-hand side of (\ref{M2}) vanishes,
\begin{eqnarray*}
[M(ak)]_{l}
&=&a(Mk)_{l}+2h^{pq}(\partial_{[p}a)k_{l]q} \\
&=&2ah^{pq}\Gamma^{s}_{q[p}k^{\hphantom{s}}_{l]s}
  +2h^{pq}(\partial_{[p}a)k_{l]q}.
\end{eqnarray*}
Substituting this in (\ref{F4}) we obtain the desired second
evolution equation:
\begin{equation}
\label{F5}
\partial_{0}f_{lmn}=\sqrt{2}[L(ak)]_{lmn}+C_{lmn},
\end{equation}
where
\begin{eqnarray*}
C_{lmn}
&=& c^{7)}_{lmn}+\sqrt{2}
  [ah^{pq}\Gamma^{s}_{q[p}k^{\hphantom{s}}_{l]s}h_{mn}
   -ah^{pq}\Gamma^{s}_{q[p}k^{\hphantom{s}}_{m]s}h_{ln} \\
& &\hspace{5mm}{}  +h^{pq}(\partial_{[p}a)k_{l]q}h_{mn}-h^{pq}(\partial_{[p}a)k_{m]q}h_{ln}].
\end{eqnarray*}

The two equations (\ref{K2}) and (\ref{F5}) constitute a first order
system for the unknowns $k_{ij}$ and $f_{lmn}$.
This system is coupled to the ordinary differential equation (\ref{ODE})
through the terms $B_{ij}$ and $C_{lmn}$ which are algebraic combinations of $h_{ij}$, $\partial_{l}h_{ij}$,
$k_{ij}$, the lapse $a$ and the shift $b$ and their spatial derivatives.
The foregoing derivation shows that if $h$ and $k$ satisfy the ADM
system (\ref{ADM1})--(\ref{ADM4}), then $h$, $k$, and $f$ satisfy
the system (\ref{ODE}), (\ref{K2}), (\ref{F5}).

\bibliography{hyper}

\end{document}